\documentclass[12pt]{article}
\pdfoutput=1

\usepackage{amsmath,amssymb,exscale}
\usepackage{array,multicol}
\usepackage{afterpage,float,flafter}
\usepackage{epsfig,rotating,pifont}
\usepackage{cite}
\@addtoreset{equation}{section} \makeatother

\def\beq{\begin{eqnarray}}
\def\eeq{\end{eqnarray}}
\def\UV{\mathrm{uv}}
\def\IR{\mathrm{ir}}
\def\Le{{\bf L}}
\def\R{{\bf R}}
\def\A{{\bf A}}
\def\I{{\rm 1\kern-.24em l}}
\usepackage{graphicx}
\usepackage{psfrag}
\usepackage{pdflscape}
\def\dd{\mathrm{d}}

\def\hc#1{{#1}^\dagger}
\def\ii{\mathit{i}}
\def\ee#1{\exp\left[{#1}\right]}
\def\tr#1{\left< #1 \right>}

\title{
\vspace{0cm}
\huge{AdS/QCD: \\ The Relevance of the Geometry}
\vspace*{0.7cm}
\author{\Large {\text{Diego Becciolini}\footnote{diego.becciolini@epfl.ch},\, \text{Michele Redi}\footnote{michele.redi@epfl.ch} \,and \text{Andrea Wulzer}\footnote{andrea.wulzer@epfl.ch}}\\ \\
\emph{ITPP, EPFL, CH-1015, Lausanne, Switzerland}}}

\date{}
\begin{document}
\maketitle \thispagestyle{empty} \vspace*{-.2cm}

\begin{abstract}
We investigate the relevance of the metric and of the geometry in five--dimensional models
of hadrons. Generically, the metric does not affect strongly the results and even flat
space agrees reasonably well with the data. Nevertheless, we observe a preference
for a decreasing warp factor, for example AdS space. The Sakai-Sugimoto model reduces
to one of these models and the level of agreement is similar to the one of flat space.
We also consider the discrete version of the five--dimensional models, obtained by
dimensional deconstruction. We find that essentially all the relevant features of
``holographic'' models of QCD can be reproduced with a simple 3-site model describing only
the states below the cut-off of the theory.
\end{abstract}


\newpage
\renewcommand{\thepage}{\arabic{page}}
\setcounter{page}{1}

\section{The Road to Five Dimensions}

The search for a dual description of QCD is a task of formidable difficulty and importance:
a success  would greatly improve the still limited understanding of quantum field theory
faraway from the perturbative regime. Such a dual model, provided it exists, should be
weakly coupled in the IR where QCD is strongly coupled and the relevant degrees of freedom are
hadrons. In the IR, therefore, the dual model should reduce to a calculable effective
(field or most likely string) theory of hadrons and one might
try, guided by the observations, to guess its broad features following a bottom--up approach.

Apart from the far IR, where the only relevant degrees of freedom are the Goldstone bosons
whose dynamics is entirely determined by the QCD global symmetries, the very existence
of such a model can be doubtful. Indeed one observes no sharp separation between the
mass and size scales of the hadrons, which seem to be all fixed by the dynamically generated
confinement scale of QCD, $\Lambda_{QCD}\sim 1$~GeV. The problem is
that no obvious weak coupling can be identified from the observational point of
view, while a weak coupling is the starting point for the construction of any
effective theory.

This difficulty immediately shows up in concrete attempts of describing hadrons,
and in particular the lightest spin--one meson, the $\rho(770)$. Both in the popular
and phenomenologically successful Hidden Local Symmetry (HLS)  approach
\cite{Bando:1984ej} (see \cite{Harada:2003jx} for a recent review) and in the
less successful (but theoretically compelling) Georgi's ``vector'' model \cite{Georgi:1989xy},
agreement with data requires a near to maximal coupling ($g_\rho\simeq6$) for the
$\rho$ meson. Even though a perturbative expansion can be set up
and loop corrections computed \cite{Harada:2003jx} by treating $g_\rho$
formally as a small expansion parameter, the fact that $g_\rho$ is numerically so
close to the perturbative bound of $4\pi$ makes us doubt that the entire
approach makes sense. Is the fact that $g_\rho<4\pi$ simply an accident
or is there a parametric reason why this happens? More precisely: is there
a deformation of the QCD theory in which $g_\rho\rightarrow0$ and treating
$g_\rho$ as a weak coupling is justified?

The answer to both questions is affirmative and has been found in the 't Hooft
large--$N_c$ limit \cite{'tHooft:1973jz}. At large--$N_c$, assuming that confinement
persists, a weak coupling emerges (see \cite{Witten:1979kh} and references therein)
and QCD becomes a weakly coupled theory of mesons. The meson masses have a
finite large--$N_c$ limit while the couplings scale as $1/\sqrt{N_c}$; this
suggests that a weakly coupled effective description of large--$N_c$
mesons must exist. It is hard to believe, however, that such a description is
provided by the models mentioned in the previous paragraph. They describe
indeed a single resonance while it is known
that large--$N_c$ mesons necessarily arise in infinite towers  \cite{Witten:1979kh}.
For those models to make sense at large--$N_c$ one should imagine,  as suggested in
\cite{Georgi:1989xy}, the higher resonances to be much heavier
than than the first one, or almost decoupled. This seems unlikely to happen.

In this context, the recently proposed 5d models of mesons
\cite{Son:2003et,Erlich:2005qh,Da Rold:2005zs,DaRold:2005vr}
(see also \cite{Polchinski:2000uf} for previous attempts)
represent a clear
progress. In these models, usually denoted as ``holographic QCD'' or ``AdS/QCD'',
the global chiral symmetry of QCD is promoted to a
local gauge symmetry in a 5d bulk and its spontaneous breaking is implemented by either a 5d
Higgs mechanism \cite{Da Rold:2005zs,DaRold:2005vr,Erlich:2005qh} or by boundary
conditions \cite{Son:2003et,Hirn:2005nr}. Both cases, which actually give very similar results, automatically deliver infinite towers of 4d
vector meson fields of fixed isospin quantum numbers. Not only this structure puts
us closer to the large--$N_c$ expectations, it also makes the model more
predictive: the ``minimal" scenario with breaking by boundary conditions only
requires 3 parameters at tree--level and describes at the same time
the $\rho(770)$, $\omega(782)$, $a_1(1260)$ and $f_1(1285)$
mesons and their towers, plus of course the pions.
\footnote{In non--minimal models, scalar resonances are also present \cite{DaRold:2005vr}.}
With the same set of parameters, the minimal model  also describes baryons which, as expected
at large--$N_c$,
are described as calculable 5d skyrmions \cite{Pomarol:2007kr,Pomarol:2008aa}.

Another important feature of the 5d construction is that it provides a simple bookkeping
of the $N_c$ factors: the $N_c$ scalings of the various observables are
recovered, both in the mesonic and baryonic sector, if taking the 5d expansion
parameter (\emph{i.e.} the gauge coupling $g_5$) to scale as $g_5\propto1/\sqrt{N_c}$.
Notice however that the quantitative agreement of the model with observation
(which can be quantified as $10\%$ in the meson sector and $30\%$ in the baryon one)
cannot be ascribed to the combined use of naive large--$N_c$ considerations and of
chiral symmetry, the 5d model gives much more precise informations. It implies for instance certain sum rules
\cite{Son:2003et,Hirn:2005nr} that we will also discuss in the following and, remarkably,
makes Vector Meson Dominance (VMD) arise automatically.

The absence of higher spin states, and of the entire Regge phenomenology, is the main
limitation of the 5d approach (see however \cite{Karch:2006pv} for a discussion
of this and related topics), even though this does not result in a concrete phenomenological
failure of the model. For the real--world case of $N_c=3$, indeed, the regime of validity
of the 5d model is limited and the observed higher spin states, which are quite heavy,
could be thought to be above the 5d cutoff. We should not worry, for the same reason,
about the masses of the heavy resonances in each tower, which will quickly become
too broad to be identified as particles and merge in the
continuum; ordinary NDA considerations suggest that it should be difficult to go above the $\rho(1450)$.
At the theoretical level, the problem shows up for larger values of $N_c$ where the coupling
decreases and the cutoff grows, but still higher spin states do not appear.
For the model to make sense as a description of large--$N_c$ we should assume higher
spins to be parametrically heavier, which seems difficult, or mostly decoupled from the vector mesons.
This seem unlikely but, according to what we know, is not in contrast with any large--$N_c$ argument.

The phenomenological road to five dimensions leaves the metric of
the 5d space completely unspecified. While originally inspired by the AdS/CFT
correspondence \cite{Aharony:1999ti}, the discussion above makes
it clear that the AdS/QCD approach can be motivated by bottom--up and
phenomenological considerations which do not require AdS geometry.
Indeed, especially at zero temperature, it appears difficult to use AdS/CFT for the strong interactions:
QCD is conformal at weak coupling while the duality becomes useful in the opposite regime, and it is
not conformal at large distances, {\it{i.e.}} in the regime relevant to determine the properties of hadrons.\footnote{See however \cite{Brodsky:2008pg} for the opposite claim.}
All this motivates us to consider five dimensional models of hadrons with general warp factor.
As long as the 4d Poincar\'e invariance is preserved, any metric would lead
to a ``reasonable'' 5d model, enjoying the general features mentioned above.
For example, the popular Sakai--Sugimoto model \cite{Sakai:2004cn} is equivalent,
 for what concerns the physics in the meson sector, to an AdS/QCD model of this kind
with a geometry different from AdS.

Since certain relations among observables will be metric--independent,
all geometries will share certain common predictions, but
the agreement with data of different models will ultimately depend on the shape of the metric.
In the rest of the paper we investigate quantitatively this dependence.
The main result is that, even though ``very wrong'' choices can be done,
a large class of metrics, including flat space, produces a good agreement
with the data. At first sight, considered the large expected corrections
to our leading order results, no metric among flat space, AdS or Sakai--Sugimoto
appears to be strongly favored. Taking the error more seriously, AdS seem to be
preferred, but its phenomenological success is shared by many other metrics, for example a warp
factor $\sqrt{L/z}$. Generically, a mildly decreasing warp factor provides a good fit of the data.
Another important feature is that the 5d models have an ``intrinsic" minimal error which depends
on the successful, but not perfect, model independent predictions.
The error is almost minimized by the AdS or $\sqrt{L/z}$ geometries.

Having quantified the dependence on the metric of the AdS/QCD results, the
natural question is whether the role of the fifth dimension, \emph{i.e.}
of the 5d geometry, is itself crucial. In fact, since the 5d theories required
to reproduce QCD are effective field theories with a limited regime of validity,
we might expect that a model with  only the few modes below the cut-off
(which turns out to be of the order 3-4) will be sufficient to reproduce similar results.
Such a model can be obtained by deconstructing the fifth dimension  \cite{ArkaniHamed:2001ca,Son:2003et}.
In the deconstructed version the five dimensional gauge theory is replaced by a 4d gauge theory
with $K$ sites and nearest neighbor interactions. We will consider models with different
number of sites and show that already the case of $K=3$
provides an good fit of the data (better than flat space) while $K=4$ is as precise as AdS.
Moreover, since in the limit $K\to \infty$ the discrete model is classically
equivalent to the 5d geometry, we use the deconstructed theory as a tool to numerically solve
arbitrarily complicated metrics.

The paper is organized as follows. In next section we review
the five-dimensional theory relevant to describe the spin 1 mesons of
QCD and its deconstructed version. In section 3 we present the results of
the fit for different low energy QCD observables for continuous and discrete
models. The validity of these effective theories is discussed in section 4.
We conclude in section 5. The appendix contains a detailed derivation
of the analog of the Chern-Simons term in the discrete models which
is necessary to reproduce the anomalies.

\section{Models of Hadrons in 5d and 4d}
\label{setup}

\subsection{Continuous Models}

We consider the same model as in \cite{Pomarol:2008aa,Grigoryan:2008cc}, a pure five-dimensional Yang--Mills theory
with $U(2)_{L} \times U(2)_{R}$ gauge group and warped extra dimension $z\in [z_{\UV}, z_{\IR}]$. We denote as $L=z_{\IR}-z_{\UV}$ the ``conformal length"
of the extra dimension and by $\Le_M$ and $\R_M$ the $U(2)_L$ and $U(2)_R$ gauge connections.
The metric on the 5d space is
\begin{equation}
 g_{MN} = a(z)^2\ \eta _{MN}\,,\label{metric}
\end{equation}
where $M$ and $N$ run over the five space-time indices, $\eta _{MN}$ is the 5d Minkowskian
metric with mostly minus signature and $a(z)$, the warp factor, is a positive and regular function.
Indexes are  raised and lowered  with the flat metric.
Any constant rescaling of the warp factor
can be reabsorbed into the definition of the 5d gauge coupling $g_5$, so that we can
normalize it to $a(z_{\IR})=1$. This chiral gauge symmetry is broken at the IR boundary by the conditions
\beq
\left(\Le_\mu-\R_\mu\right)\left|_{z=z_{\IR}}\right.=0\ ,\;\;\;\;\;\;\left(\Le_{\mu 5}+\R_{\mu 5}\right)\left|_{z=z_{\IR}}\right.=0\, ,
\label{irboundary condition}
\eeq
where the 5d field strength is $\Le_{MN}=\partial_M \Le_N-\partial_N \Le_M-i[\Le_M,\,\Le_N]$
and analogously for $\R_{MN}$. At the UV boundary the fields can be identified with the external sources
for the QCD global currents
\beq
 \Le_\mu\left|_{z=z_{\UV}}\right.=\,{\bf l}_\mu\ , \;\;\;\;\;\; \R_\mu\left|_{z=z_{\UV}}\right.=\,\bf {r}_\mu \, .
\label{uvboundary condition}
\eeq
We will consider general functional forms of $a(z)$, including for instance flat space, $a(z)=1$.
The geometry, therefore, does not distinguish between the UV and IR boundaries, and what
makes them different are the boundary conditions (\ref{irboundary condition},\ref{uvboundary condition}).

Thanks to the IR boundary conditions (\ref{irboundary condition}) we can reformulate our model,
which describes two $U(2)$ gauge fields $\Le$ and $\R$ on the space $z\in [z_{\UV}, z_{\IR}]$,
in terms of a single  $U(2)$ field $\A$ living on a ``doubled'' 5d space with extra coordinate
$u\in [-L,L]$. This is done by ``gluing'' $\Le$ and $\R$ together in the following way:
\beq
 \A_\mu(u) &\equiv& \left\{
 \begin{array}{l}
  \Le_\mu( z_{\IR} + u),\quad u \in [-L,0] \\
  \R_\mu( z_{\IR} - u),\quad u \in [0,L]
 \end{array} \right. \,,\nonumber \\
 \A_5(u) &\equiv& \left\{
 \begin{array}{l}
  \Le_5( z_{\IR} + u),\quad u \in [-L,0] \\
  - \R_5( z_{\IR} - u),\quad u \in [0,L]
 \end{array} \right.  \,.
\label{glue}
\eeq
Notice that the IR boundary conditions (\ref{irboundary condition}), which $\Le$ and
$\R$ have to respect, ensure that the field $\A$ is continuous at $u=0$.

The doubled space has two boundaries $u=-L$ and $u=L$ which we call respectively the
left and right boundaries. The boundary conditions for $\A$, obtained by combining
eq.s~(\ref{uvboundary condition}) and (\ref{glue}), are
\begin{equation}
 \A_\mu(-L) = {\bf l}_\mu,\qquad
 \A_\mu(L) = {\bf r}_\mu\,.
\label{uvbc}
\end{equation}
The metric of the doubled space is of the form (\ref{metric}) with a warp factor
\beq
a(u) &\equiv& \left\{
 \begin{array}{l}
  a(z_\IR + u),\quad u \in [-L,0] \\
  a(z_\IR - u),\quad u \in [0,L]
 \end{array} \right.\,
\eeq
which is symmetric under the reflection $u\rightarrow-u$.
Being this transformation an isometry, we could impose it as a symmetry of our model which
however would not correspond to any of the QCD symmetries.
\footnote{This would rather correspond to the $U\rightarrow U^\dagger$ ``accidental''
symmetry of the ordinary pion Lagrangian, which is broken by the Wess-Zumino-Witten term. Similarly
(see below) $u\rightarrow-u$ is broken by the CS term.}
 What corresponds to the QCD
parity is, in this language, the simultaneous inversion of all spacial coordinates, \emph{i.e.} the
combined action of $u\rightarrow-u$ and $\vec{x}\rightarrow-\vec{x}$ where $\vec{x}$ denotes
3--space coordinates. In the language of $\Le$ and $\R$ fields this symmetry corresponds, via eq.~(\ref{glue}), to the $L\leftrightarrow R$ interchange combined with $\vec{x}\rightarrow-\vec{x}$.

Finally, the action reads
\beq
 S_\mathrm{YM} &=& -\frac{1}{2 {g_5}^2} \int \dd ^5 x\ a(u) \left[ \tr{F_{MN} F^{MN}} + \frac {\alpha^2} {2}\left({\widehat{F}_{MN} \widehat{F}^{MN}}\right) \right] \label{actionym}\\
 S_\mathrm{CS} &=& \frac{N_c}{16\pi^2}\int \dd^5x \left[ \frac{1}{4} \epsilon^{MNOPQ} \widehat{A}_M \tr{F_{NO}F_{PQ}} + \frac{1}{24} \epsilon^{MNOPQ} \widehat{A}_M \widehat{F}_{NO} \widehat{F}_{PQ} \right]
\label{action}
\eeq
where the field strength is $F_{MN} \equiv \partial_M A_N - \partial_N A_M - \ii \left[ A_M, A_N \right]$,
 we have parametrized the $U(2)$ gauge field $\A$ as $\A_M=A_M^a\sigma_a/2+
\widehat{A}_M\I/2$ in terms of Pauli matrices and $\tr{\, .\, }$ denotes the trace.
In the previous action we introduced different couplings for $U(1)$ and $SU(2)$ sectors
as this is allowed by the symmetries of the model.  However it is consistent
with large $N_c$ to set them equal, so we will work with $\alpha=1$ in the rest of the paper for convenience,
 since it would not affect the results much.
The presence of the Chern-Simons term $S_{CS}$ is  required for matching the anomalies of
large--$N_c$ QCD and its coefficient is fixed
by the number of colors $N_c$. For a given warp factor $a(z)$, therefore, the model has $2$ free
parameters to be determined by experiments.
\footnote{
AdS/QCD is often regarded as a 1--parameter model since the $g_5$ coupling is
``matched'' at the UV to the perturbative QCD result \cite{Erlich:2005qh,DaRold:2005zs}.
In the spirit of the introduction we instead consider $g_5$ as a purely IR parameter and regard
as a mere coincidence the fact that its matched value would be close, in the case of AdS,
to the phenomenologically preferred one.}
Notice that the simplified form of the CS that we use
is only valid in the two flavors case, \emph{i.e.} for $U(2)$ 5d gauge fields. It is very simple,
splitting the $u$ domain of integration into $u \in [-L,0] $ and $u \in [0,L] $ and using
eq.~(\ref{glue}), to show that the action (\ref{action})  is equal to the one considered in
\cite{Pomarol:2008aa} once expressed in terms of $\Le$ and $\R$ fields. The two theories
are therefore equivalent up to the choice of the warp factor which was fixed to $a(z)=L/z$
in \cite{Pomarol:2008aa} while we will consider more general possibilities.

One of the advantages of rewriting the model on the doubled space is that the connection
with the Sakai-Sugimoto model becomes manifest.
This model is motivated by strings but effectively reduces, in the limit of large string scale,
to a gauge field theory on warped 5d space of the kind considered in this paper. All the
calculations of mesons properties \cite{Sakai:2004cn} are performed in the
field theory regime, and are insensitive to the string completion. The action is
\begin{equation}
S_\mathrm{YM} = -\frac{1}{2 {g_5}^2} \int \dd ^4 x dz\  \left[
K(z)^{-\frac 1 3}\tr{F_{\mu\nu} F^{\mu\nu}}+2\,K(z)\,\tr{F_{\mu z} F^{\mu z}}\right]\,,
\end{equation}
with $K= 1+z^2$ plus the same Chern-Simons term as in (\ref{action}).
By the change of coordinates
\begin{equation}
\frac {d z}{du}=(1+z^2)^{\frac 2 3}\,,
\end{equation}
(and rescaling $u$ suitably) the action above can be rewritten in our frame with warp factor
\begin{equation}
a(u)=K(z(u))^{\frac 1 3}\,.
\label{ssmetric}
\end{equation}
Therefore the Sakai--Sugimoto model is at the practical level equivalent to an AdS/QCD
 model of the kind we consider here.

It would be straightforward, for a generic warp factor, to study the
phenomenological consequences of the setup described in this section. It is only
for few particular choices of the warp factor, however, that the model
could be solved analytically. For a generic warp factor we should in any case
rely on numerical methods to solve the differential equations which determine
the masses and wave functions of the KK resonances. This motivates us to study
the continuous 5d models as a limiting case of the discrete model to which we now turn.
The problem of finding wave functions and masses will be reduced to the one of
diagonalizing ``large'' matrices, a numerically easier task.
The discrete model, however, is not just a technical tool. It can be considered
per s\`e, for a limited number of sites, as a model of hadrons. It is only
for an infinite number of sites that it (classically) coincides with the 5d model.

\subsection{Discrete Models}

\begin{figure}[t]
\
\vspace{-1.5cm}
\begin{center}
\epsfig{file=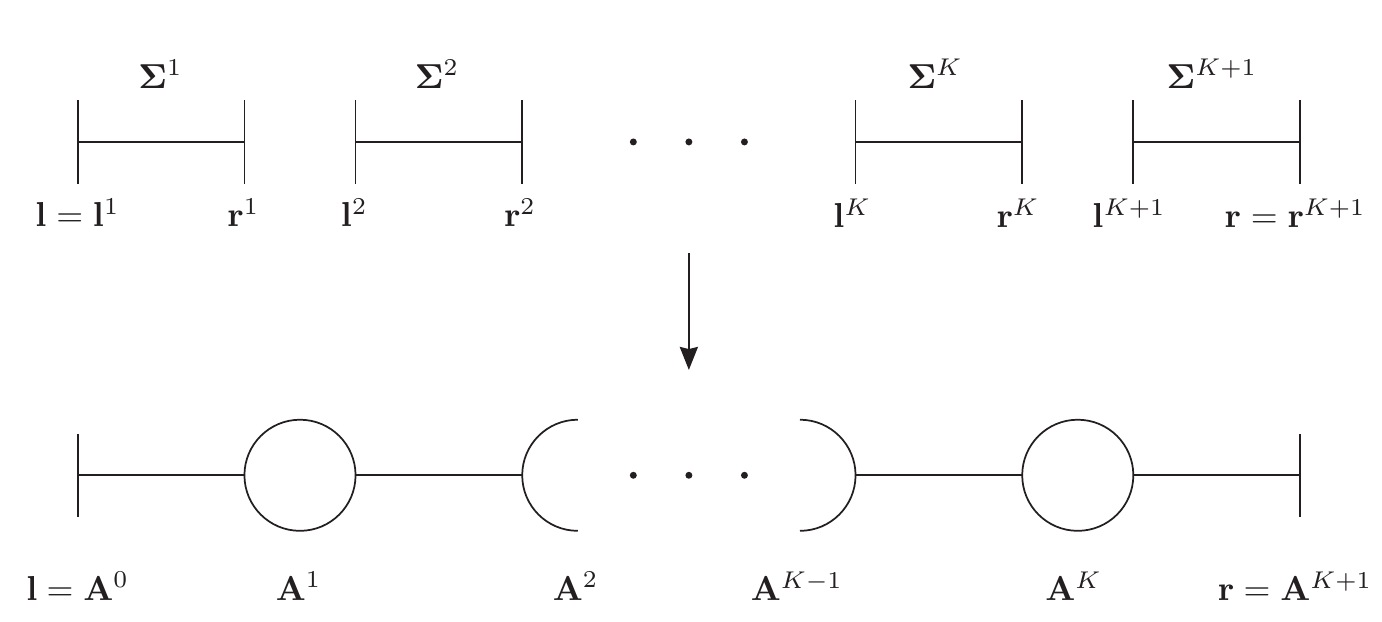,width=0.8\textwidth,angle=0}
\end{center}
\caption{
A graphical representation of the discrete model.}
\label{decfig}
\end{figure}

What we denote as discrete models can be considered to be generalizations of Georgi's ``vector'' model
\cite{Georgi:1989xy} to the case of many sites, \emph{i.e.} of many non--linearly realized gauge groups.
The models we are considering are therefore the ones discussed in
ref.~\cite{Son:2003et}, but with $U(2)$ instead of $SU(2)$ groups and
the addition of a gauged Wess--Zumino--Witten (WZW) term at
each link. The latter terms are necessary to reproduce the QCD anomaly and provide,
expectedly, a discretized version  of the 5d CS term  \cite{Skiba:2002nx}.

A diagrammatic representation of the discrete model is shown in
figure~\ref{decfig}: each
of the $K+1$ links represents a $U(2)$ $\sigma$--model matrix ${\bf\Sigma}^k$
($k=1,\ldots,K+1$) transforming as
\beq
{\bf\Sigma}^k\;\rightarrow\;{\bf L}_{k}{\bf\Sigma}^k{\bf R}_{k}^\dagger\,,
\label{ggroup}
\eeq
under a $\left(U(2)_L\times U(2)_R\right)_{k}$ global group of elements
${\bf L}_{k}$ and ${\bf R}_{k}$.
The associated non--dynamical sources are denoted as ${\bf l}^{k}$ and ${\bf r}^{k}$
in figure~\ref{decfig}. For the time being, different $\sigma$--models do not interact with each other
and at two derivatives order the Lagrangian allowed by the symmetries (\ref{ggroup})
is
\beq
\sum_{k=1}^{K+1}\frac{f_k^2}{4}\,\left[
 \tr{\left|\partial_\mu \Sigma^k \right|^2} +
\frac{1}{2} \left|\partial_\mu {\widehat\Sigma}^k \right|^2\right]\,,
\label{lag_ung}
\eeq
where we have separated each $U(2)$ matrix ${\bf\Sigma}^k={\widehat\Sigma}^k \Sigma^k$
in its $SU(2)$ ($\Sigma^k$) and $U(1)$ (${\widehat\Sigma}^k$) components. As in eq. (\ref{action})
we choose the parameters of the $U(1)$ and $SU(2)$ sectors to be equal. Notice that we
could have enlarged the symmetry group of the theory if we had chosen all the $f_k$
to be equal in eq.~(\ref{lag_ung}). With this choice we would have gained the discrete
symmetry of links permutation and greatly reduced the number of parameters.

Let us now make different links interact by performing a weak gauging of a subgroup
of the $\left(U(2)_L\times U(2)_R\right)^{K+1}$ global group;
this partially breaks the global symmetry which forced us to write the Lagrangian
(\ref{lag_ung}), but in a controlled way, as explained in \cite{Georgi:1989xy}.
The groups we want to gauge are the vector combination of the Right group of each
link ({\it i.e}, ${\bf R}_{k}$ in eq.~(\ref{ggroup})) with the Left group of the following
one ({\it i.e}, ${\bf L}_{k+1}$). The gauging procedure is depicted in the lower part of
figure~\ref{decfig}; each gauged group is denoted by a circle and the
associated gauge fields by ${\bf A}_\mu^k$.
The gauging consists in introducing the covariant derivatives
\beq
D_\mu {\bf\Sigma}^k &\equiv& \partial_\mu {\bf\Sigma}^k - \ii \A^{k-1}_\mu {\bf\Sigma}^k + \ii
{\bf\Sigma}^k \A^k_\mu \,,
\label{cder}
\eeq
and adding a kinetic term to the ${\bf A}^k$, the Lagrangian reads
\beq
{\mathcal L}_g=
\sum_{k=1}^{K+1}\frac{f_k^2}{4}\,\left[
  \tr{\left|D_\mu \Sigma^k \right|^2} +
 \frac{1}{2} \left|D_\mu {\widehat\Sigma}^k \right|^2\right]
 -
 \sum_{k=1}^{K}\frac{1}{2 {g_k}^2}\,\left[
   \tr{\left(F^k_{\mu\nu}\right)^2}
 + \frac{1}{2}
 \left(\widehat{F}^k_{\mu\nu}\right)^2
 \right]
 \,.
 \label{lag}
\eeq

The QCD Left and Right chiral transformations are identified, respectively,
with ${\bf L}_1$ and ${\bf R}_{K+1}$ in eq.~(\ref{ggroup}), which are left
ungauged. The fields $\A^0$ and $\A^{K+1}$ appearing in the covariant
derivative of, respectively,
${\bf\Sigma}^0$ and ${\bf\Sigma}^{K+1}$ are defined as the Left ($\A^0={\bf l}$) and
Right ($\A^{K+1}={\bf r}$) QCD sources and are therefore non--dynamical field.
They couple to the currents,
\begin{equation}
 {J_L}_\mu^a = \ii \frac{f_1^2}{4} \left. \left((D_\mu\Sigma^1)\hc{(\Sigma^1)}\right)^a \right|_{\mathit{l}_\mu=0},\qquad {J_R}_\mu^a = \ii \frac{f_{K+1}^2}{4} \left. \left(\hc{(D_\mu\Sigma^{K+1})}(\Sigma^{K+1}) \right)^a \right|_{\mathit{r}_\mu=0}\,,
\label{currents}
\end{equation}
which we identify with the $SU(2)$ QCD currents and analogous expressions for the $U(1)$ currents.

In this model QCD anomalies can be reproduced by introducing WZW terms at each site \cite{Skiba:2002nx}.
We postpone the full derivation of this term to the appendix. The final result is simply,
\beq
S_{A} &=& \ii \frac{N_c}{96\pi^2} \int \sum_{k=1}^{K+1}
\Biggl[ \widehat{A}^{k-1}\, \widehat{A}^k\, ( \dd\widehat{A}^{k-1} + \dd\widehat{A}^{k} )
- \ii{\widehat\Pi}^{k}
\left( (\dd \widehat{A}^{k-1})^2 + (\dd \widehat{A}^{k})^2 +
 \dd \widehat{A}^{k-1}\, \dd \widehat{A}^{k} \right)
\notag\\&&- 3\ii\, {\widehat\Pi}^k \tr{(F^{k-1})^2 + (F^k)^2}
 +\ 3\, (\widehat{A}^{k-1} + \widehat{A}^{k})
\tr{ F^{k-1} \Sigma^k D\hc{\Sigma^k} - F^k \hc{\Sigma^k} D\Sigma^k }
 \notag\\&&+ (\widehat{A}^{k-1} + \widehat{A}^{k})
\tr{\Bigl( \hc{\Sigma^k} D\Sigma^k \Bigr)^3 } \Biggr]\,, \label{cs}
\eeq
where ${\widehat\Pi}^{k}$ is defined by ${\widehat\Sigma}^{k}=\exp{i\,{\widehat\Pi}^{k}}$ and we used a form notation in which $A=-\ii  A_\mu dx^\mu$ and $F=dA+A^2$. Notice the addition of $S_A$ to the action makes an extra contribution to the $U(1)$ currents (\ref{currents}) arise. The latter can however be neglected since it will not contribute to the observables we will be interested in.

This completes the construction of the discrete model. We still need to impose
invariance under the QCD parity operation, which we identify with the reflection with
respect to the middle point
of the diagram in figure~\ref{decfig}, \emph{i.e.} as
${\bf\Sigma}^k\leftrightarrow \hc{({\bf{\Sigma}}^{{K+2-k}})}$ and $\A^k\leftrightarrow\A^{K+1-k}$
combined of course with ordinary ${\vec{x}}\rightarrow-{\vec{x}}$.
The reason for this identification is that parity must  interchange  the Right source
${\bf r}=\A^{K+1}$ with the Left one ${\bf l}=\A^0$. In order for the action
to be invariant we have to take $f_{k}=f_{K + 2 -k}$ and $g_{k}=g_{K+1-k}$.

It is important to remark that the logic we have followed in the formulation of the model led us to
write a Lagrangian of special ``nearest--neighbor" form
\beq
{\mathcal L}=\sum_{k=1}^{K+1}{\mathcal L}_k\left({\bf\Sigma}^k,\A^{k-1},\A^{k}\right)\,.
\label{nnl}
\eeq
Non-nearest neighbor interactions will be radiatively generated by the gauge fields
dynamics which breaks the symmetry (\ref{ggroup}) and makes different links interact.
Their coefficients, however, will be suppressed by powers
of the gauge couplings $g_k$, which are the parameters
controlling the breaking of the symmetry (\ref{ggroup}). In an expansion in the
gauge couplings, therefore, non--nearest neighbor operators are subleading
and we are allowed not to consider them at the leading order.
The situation is exactly the same as in the model of ref.~\cite{Georgi:1989xy}
which corresponds to the $K=1$ case. It is clear from the discussion above
that the discrete model is based on a weak coupling (small $g_k$)
expansion, and that it is not meaningful to consider gauge couplings which exceed $4\pi$.
We will however use formally infinite values of the couplings when we will take the
``continuous limit'' of the discrete model: we have already mentioned that this limiting
procedure can only be thought as a numerically convenient method to obtain predictions in
the 5d theory with general metric.

Following \cite{Son:2003et,ArkaniHamed:2001ca}, the discrete model becomes equivalent to the
continuous one described in the previous section when the the number of sites $K$ is taken to infinity
while keeping ${g_k}^2/K$ and ${f_k}^2/K$ fixed. The relation between 4d and 5d models is
given by
\beq
\displaystyle
g_k^2 \,=\, \frac{{g_5}^2}{a(u_k)}\frac{K+1}{2L}\,,\qquad
f_k^2 \,=\, \frac{a\left[(u_{k-1}+u_k)/2\right]}{{g_5}^2}\frac{2\,(K+1)}{L}\,.
\label{dl}
\eeq
It is already clear that we will identify the parameters $L$ and $g_5$
and the function $a(u)$ with the ones appearing in the continuos 5d model.
Moreover, the $u_k$  are points on the discretized extra--dimension $u$,
whose values are
\beq
u_k\,=\,u_0\,+\,\frac{2L}{K+1}\,k\,,
\eeq
with $k=0,\ldots,K+1$ and $u_0=-L$.

In the continuous limit, the fields ${\bf \Sigma}^k$ are identified with the link variables
of lattice gauge theories, \emph{i.e.} with the Wilson lines on a straight path from the $u_{k}$
to the $u_{k-1}$ points, while $\A_\mu^k$ is the 5d gauge field at the point $u_k$. It
follows that, in this limit
\footnote{What had to be shown is that the only field configurations
relevant in the continuous limit are in the form of eq.~(\ref{clf})
where, for instance, ${\bf\Sigma}^k$ only deviates from the identity by
an ${\mathcal O}(1/K)$ term. This could be checked explicitely to hold for the
eigenmodes that we will derive in the following section.}
\begin{eqnarray}
&&
{\bf\Sigma}^k\,\simeq\,\I\,-\,\ii\,\frac{2L}{K+1}\,{\A}_5\left[(u_{k-1}+u_k)/2\right]\,,
\qquad \A_\mu^k\,\simeq\,\A_\mu\left[u_k\right]\,,\nonumber\\
&&
\left(\A_\mu^k-\A_\mu^{k-1}\right)\,\simeq\,\frac{2L}{K+1}\,\left(\partial_5\A_\mu\right)
\left[(u_{k-1}+u_k)/2\right]\,.
\label{clf}
\end{eqnarray}
It is simple to check, using the equations above and changing the sum into
an integral, that the Lagrangian (\ref{lag}) reduces to the 5d $S_{YM}$
action in eq.~(\ref{actionym}). Similarly $S_{A}$ reconstructs,
the 5d CS action of eq.~(\ref{action}). The first line of eq. \ref{cs} reduces, using
eq.~(\ref{clf}), to the second term of eq.~(\ref{action}). One has to use that
$\widehat{A}^{k-1}\widehat{A}^k=(\widehat{A}^{k-1}-\widehat{A}^k)\widehat{A}^k$
and neglect terms which go to zero faster than $1/K$. For the second line, since
$\hc{(\Sigma^k)}D_\mu\Sigma^k\simeq-\ii 2L/(K+1)F_{\mu 5}$ one obtains
the first term of eq.~(\ref{action}). Moreover, the third line vanishes in the continuous
limit as it scales like $1/K^3$.

\subsection{QCD Observables}
\label{observables}

We now present the expressions for the QCD observables to be compared with the data.
We will derive the formul\ae\ in the discrete model; the analogous expressions in the continuous
limit can be quickly obtained by taking $K\to \infty$ and using eqs. \eqref{dl}.

\subsubsection{Chiral Lagrangian}

The effective Lagrangian for the pions up to fourth order in the derivatives is customarily
parametrized as follows \cite{Gasser:1984gg}
\beq
 \mathcal{L}_{p^2} &=& \frac{f_\pi^2}{4} \tr{\hc{(D_\mu \Sigma)} (D^\mu \Sigma)}\,, \\
 \mathcal{L}_{p^4} &=&
 L_1 \tr{\hc{(D_\mu \Sigma)} (D^\mu \Sigma)}^2
 + L_2 \tr{\hc{(D_\mu \Sigma)} (D_\nu \Sigma)} \tr{\hc{(D^\mu \Sigma)} (D^\nu \Sigma)} \\
 && +\ L_3 \tr{\hc{(D_\mu \Sigma)} (D^\mu \Sigma) \hc{(D_\nu \Sigma)} (D^\nu \Sigma)} \notag\\
 && -\ \ii L_9 \tr{\mathit{l}_{\mu\nu}(D^\mu\Sigma)\hc{(D^\nu\Sigma)} +
 \mathit{r}_{\mu\nu}\hc{(D^\mu\Sigma)}(D^\nu\Sigma) }
 + L_{10} \tr{\hc{\Sigma}\mathit{l}_{\mu\nu}\Sigma\mathit{r}^{\mu\nu}}\,, \notag
\eeq
where the covariant derivative is $D_\mu\Sigma \equiv \partial_\mu\Sigma - \ii\,\mathit{l}_\mu\Sigma + \ii\,\Sigma\mathit{r}_\mu$.
The one above is the purely ``$SU(2)$'' part of the Goldstone Bosons Lagrangian, and it will not receive contributions from the
 $S_A$ term in eq.~(\ref{cs}). We therefore ignore the latter in the discussion which follows.

To extract the coefficients of the pion lagrangian from our action it is convenient to follow the ``holographic''
approach of ref.~ \cite{Panico:2007qd}, appropriately adapted to the discrete case. To do this we first choose
a gauge where all the link variables $\Sigma^k$ but the first one, denoted as $\Sigma$, are set to the identity
 and then
notice that the dependence on $\Sigma$ of the gauge--fixed action can be reabsorbed in the redefinition of
$A_\mu^0$. Namely, the action can be written as in eq.~(\ref{lag}) with all the $\Sigma^k$'s set to the
identity but with
\begin{equation}
 A_\mu^0 = \hc{\Sigma}(\mathit{l}_\mu + \ii \partial_\mu)\Sigma\,,\qquad A_\mu^{K+1} = \mathit{r}_\mu\,.
\end{equation}
In terms of the original fields (before fixing the gauge), $\Sigma$ is
\begin{equation}
\Sigma = \Sigma^1 \Sigma^2 \cdots \Sigma^{K+1}\,,
\end{equation}
and represents  the discretized version of the Wilson line  from one boundary to the other.

We will now write the effective theory for the $\Sigma$ field,  which describes the physical pion, by integrating out the gauge fields $A_\mu^k$ at tree--level, {\it {i.e.}} by solving the classical equations of motion and plugging back
into the action. It is possible to check that, like in the continuous case of \cite{Panico:2007qd}, it is enough to
consider the linearized equation at zero momentum in order to obtain the effective Lagrangian up to $\mathcal{O}(p^4)$.
The soluton is
\begin{equation}
A_\mu^k = \frac{1}{2}\left(1 + b_0^k\right)\hc{\Sigma}(\mathit{l}_\mu + \ii \partial_\mu)\Sigma + \frac{1}{2}\left(1 - b_0^k\right)\mathit{r}_\mu\,,
\end{equation}
where
\begin{equation}
b_0^k = 1 - 2 \left(\sum_{i=1}^{K+1}{\frac{1}{f_i^2}}\right)^{-1} \left(
 \sum_{i=1}^{k}{\frac{1}{f_i^2}} \right)\,.
\end{equation}

The $\mathcal{O}(p^2)$ and $\mathcal{O}(p^4)$  terms in the chiral Lagrangian arise, respectively, from the first
and second term of the Lagrangian \eqref{lag}. We immediatly read the pion decay constant
\begin{equation}
\frac{1}{f_\pi^2} = \sum_{k=1}^{K+1} \frac{1}{f_k^2}\,,
\label{fpidiscr}
\end{equation}
which in the continuos limit this reduces to the standard,
\begin{equation}
f_\pi^2 = \frac{2}{g_5^2} \left( \int_{0}^{L} \frac{\dd u}{a(u)} \right)^{-1}\,.
\label{fpicont}
\end{equation}
For the order $\mathcal{O}(p^4)$ terms one finds,
\begin{eqnarray}
L_1 = \frac{1}{2} L_2 = -\frac{1}{6} L_3 &=& \frac{1}{32}\sum_{k=1}^K \frac{1}{g_k^2} \left(1-{(b_0^k)}^2\right)^2\,, \label{L1} \nonumber \\
L_9 = -L_{10} &=& \frac{1}{4}\sum_{k=1}^K \frac{1}{g_k^2} \left(1-{(b_0^k)}^2\right)\,, \label{L9}
\end{eqnarray}
which again reduce to the standard 5d formul\ae\ in the continuous limit.

\subsubsection{Vector Mesons}

The other observables are more easily obtained performing the analog of the Kaluza-Klein reduction.
To do this it is convenient to proceed as in ref.~\cite{Son:2003et}. By choosing the gauge
\begin{equation}
\Sigma^k = \ee{2\ii\frac{f_\pi}{f_k^2}\pi}\,,
\label{sigma}
\end{equation}
the scalars do not mix with the vector fields. We can then diagonalize the mass
matrix for the massive gauge bosons. To do this we introduce
\begin{equation}
A_\mu^k(x)=g_k\, \sum_{n=1}^K b_n^k\, B_{\mu}^n(x)\,,
\end{equation}
where $B_{\mu}^n(x)$ is a vector field with mass $m_n$.

Masses and wave functions are determined by the following eigenvalue problem:
\begin{equation}
g_k\, \left( f_k^2\; g_{k-1}\; b^{k-1}_n - (f_k^2 + f_{k+1}^2)\; g_k\; b^k_n + f_{k+1}^2\; g_{k+1}\; b^{k+1}_n \right) + 4\, m_n^2\; b^k_n = 0\,,
\label{kkmodes}
\end{equation}
for $k\in[1,K]$ and where $b_n^0 = b_n^{K+1} = 0$, since the $l$ and $r$ sources are non-dynamical and have to be put to zero.

Due to the symmetric nature of our action the modes split into even ($b_n^k = b_n^{K+1-k}$) and odd ($b_n^k = -b_n^{K+1-k}$) whose masses alternate in the spectrum. The actual solutions (with unit norm) can be easily found numerically for given $f_k$ and $g_k$ (in particular if we fix them with the relation \eqref{dl} for a given metric).

The trilinear vertices we are interested in are
\begin{eqnarray}
\mathcal{L}_{B^n\pi\pi} &=& g_{B^n\pi\pi}\cdot \epsilon^{abc}\ {B_\mu^n}^{\,a}\, \pi^b\, \partial^\mu\pi^c\,,\nonumber  \\
\mathcal{L}_{\hat{B}^m B^n\pi} &=& -g_{\hat{B}^m B^n\pi}\cdot \epsilon^{\mu\nu\rho\sigma}\ \partial_\mu \hat{B}^m_\nu\; \partial_\rho {B_\sigma^n}^{\,a}\; \pi^a\,,\nonumber \\
\mathcal{L}_{B^n\pi\gamma} &=& -g_{B^n\pi\gamma}\cdot \epsilon^{\mu\nu\rho\sigma}\ \partial_\mu B^\gamma_\nu\; \partial_\rho {B_\sigma^n}^{\,a}\; \pi^a\,, \nonumber \\
\mathcal{L}_{\hat{B}^n\pi\gamma} &=& -g_{\hat{B}^n\pi\gamma}\cdot \epsilon^{\mu\nu\rho\sigma}\ \partial_\mu B^\gamma_\nu\; \partial_\rho \hat{B}_\sigma^n\; \pi^3\,.
\end{eqnarray}

The first two couplings can be obtained by simply plugging the wave-functions (\ref{kkmodes}) into the action.
To obtain the last two coupling we need to gauge the electromagnetism. In the language of the previous section
the electromagnetic current corresponds to the combination $J^\mu_{EM} = 1/3\ {\widehat{J}}_V^\mu + {J^3_V}^\mu$.
To make the photon dynamical (see \cite{Pomarol:2008aa} for the discussion in the continuous case)
we just need to add a kinetic term for the source associated to the current, \emph{i.e.} for the appropriate components of $A_0$ and $A_{K+1}$. We could now repeat the full Kaluza-Klein reduction. However it is much easier to recognize
that to leading order in the electric charge the solution is simply
\begin{eqnarray}
{A^k_\mu}^{\,a}(x) &=& e\, B^\gamma_\mu(x)\ \delta^{\,a3} + \sum_{n=1}^K\ g_k\ b^k_n\ {B^n_\mu}^{\,a}(x) \label{KKdiscr1}\,,\nonumber  \\
\widehat{A}^k_\mu(x) &=& \frac{1}{3}\ e\, B^\gamma_\mu(x) + \sum_{n=1}^K\ g_k\ b^k_n\ \widehat{B}^n_\mu(x)\,,
\label{KKdiscr2}
\end{eqnarray}
which follows from the fact that the heavy modes are unchanged to leading order in the perturbation.
Using these wave-functions we obtain
\begin{eqnarray}
g_{B^n\pi\pi} &=& \frac{{f_\pi}^2}{2} \sum_{k=1}^{K+1} \frac{1}{f_k^2}\, \Bigl(g_{k-1}\, b_n^{k-1} + g_k\, b_n^k \Bigr)\,,\nonumber \\
 g_{\hat{B}^m B^n\pi} &=& \frac{N_c\, f_\pi}{32\pi^2}\sum_{k=1}^{K+1} \frac{1}{f_k^2}\, \Bigl( g_{k-1}\, b_m^{k-1} + g_k\, b_m^k \Bigr) \Bigl( g_{k-1}\, b_n^{k-1} + g_k\, b_n^k \Bigr)\,,\nonumber\\
g_{B^n\pi\gamma} &=& \frac{e\, N_c}{24\pi^2 f_\pi}\ g_{B^n\pi\pi} \label{su2pionphoton}\,,\nonumber \\
g_{\hat{B}^n\pi\gamma} &=& \frac{e\, N_c}{8\pi^2 f_\pi}\ g_{B^n\pi\pi}\,. \label{u1pionphoton}
\end{eqnarray}
Notice that the second coupling is non zero only if $b_m^k$ and $b_n^k$ have the same parity while the others only if $b_n^k$ is even under the reflection \mbox{$k\leftrightarrow K+1-k$} (\emph{i.e.} for vector resonances).

The last observables that we consider are decay constants (for axial and vectors mesons) and the pion radius.
As for $f_\pi$, the meson decay constants  are defined in terms
of the matrix elements of the QCD currents between the vacuum and the one meson particle states
\begin{equation}
\left< 0 \,|\, J^a_\mu \,|\, n^b \right> = -\delta^{ab} \epsilon_\mu m_n F_n\,.
\end{equation}
By using the KK decomposition and the currents (\ref{currents}) one finds,
\begin{equation}
F_n = \frac{f_1^2\, g_1}{4\,m_n} \left( b_n^1 \pm b_n^K \right) = \frac{f_1^2\, g_1}{2\,m_n}\, b_n^1\,,
\end{equation}
where $+$ corresponds to the vector decay constant and $-$ for the axial (obviously
the only non zero matrix elements are between vector (axial) currents and vector (axial) mesons).

Finally the pion radius is defined in terms of the pion form factor as
\begin{equation}
\left< r_\pi^2 \right> = 6 \left. \frac{\partial \mathcal{F}_\pi(q^2)}{\partial q^2} \right|_{q^2=0}\,,
\end{equation}
where
\begin{equation}
\left<\pi^a(p')\,| J^b_\mu(0)\,|\pi^c(p)\right>=i\,\epsilon^{abc}(p+p')_\mu\, {\cal F}_\pi(q^2) \,.
\end{equation}
with $q_\mu=p_\mu-p'_\mu$. The pion form factor can be expressed as,
\begin{equation}
\mathcal{F}_\pi(q^2) = \frac{f_\pi^2}{f_1^2} + \sum_{n=1}^K \frac{m_n\, F_n\; g_{B^n\pi\pi}}{m_n^2 - q^2}\,,
\label{pionformfactor}
\end{equation}
from which one finds that,
\begin{equation}
\left< r_\pi^2 \right> = 6\, \sum_{n=1}^K \frac{ F_n\; g_{B^n\pi\pi}}{m_n^3 }\,.
\label{pionradius}
\end{equation}
Actually a sum rule relates $r_\pi^2$ to $L_9$ and $f_\pi$ (see eq. \eqref{L9sum})
so that $\left< r_\pi^2 \right> = 12 L_9/f_\pi^2$.

\subsubsection{Sum Rules}

The discrete models satisfy several sum rules \cite{Son:2003et} (see \cite{Hirn:2005nr} for the continuous case),
 all of which can be proven using the definitions given in the previous sections, the wave-equation \eqref{kkmodes} and the orthogonality of the eigenmodes.

For example (generalized) Weinberg sum rules
\beq
\sum_{n=1}^K\ (-1)^{n+1}\ \frac{F_{n}^2}{m_n^2} &=& 4 L_9\,, \nonumber  \\
\sum_{n=1}^K\ (-1)^{n+1}\ F_{n}^2 &=& f_\pi^2\,, \nonumber  \\
\sum_{n=1}^K\ (-1)^{n+1}\ m_{n}^{2j}\, F_{n}^2 &=& 0, \qquad j\in[1,K-1]\,,
\eeq
and the following relations
\beq
L_1 &=& \frac{f_\pi^4}{8}\; \sum_{n=1}^K \frac{g_{B^n\pi\pi}^2}{m_n^4}\,, \label{L1sum} \nonumber \\
L_9 &=& \frac{f_\pi^2}{2}\; \sum_{n=1}^K \frac{F_n\; g_{B^n\pi\pi}}{m_n^3}\,, \label{L9sum}
\eeq
hold independently of the number of sites.

Other sum rules however receive corrections for finite $K$. In particular
vector meson dominance (VMD) is exact only in the continuous limit. From the fact that $\mathcal{F}_\pi(0) = 1$, form eq.
(\ref{pionformfactor}) one has
\begin{equation}
\sum_{n=1}^K\ \frac{F_n\; g_{B^n\pi\pi}}{m_n} = 1 - \left( \frac{f_\pi}{f_1} \right)^2\,.
\label{VMD}
\end{equation}
To reproduce VMD $f_1$ should be significatively larger than $f_\pi$.
As we will see this trend is reproduced by our fit of the data.

The other sum rule which will be relevant in our phenomenological study is given by
\begin{equation}
f_\pi^2\; \sum_{n=1}^K \frac{g_{B^n\pi\pi}^2}{m_n^2}= \frac 1 3 -\frac 1 3\, \sum_{k=1}^{K+1} \left( \frac{f_\pi}{f_k}\right)^6\,.  \label{fpisum1}
\end{equation}
This equation implies a deviation from the so called KSRF formula \cite{KSRF}, the phenomenological relation
$g_{\rho\pi\pi}^2 f_\pi^2/m_\rho^2 = 1/2$, which is well verified experimentally.
From the equation above it follows that this ratio is always less than $1/3$.
In comparison to continuous models, for finite number of sites there
is an extra negative contribution to this relation which however quickly goes
to zero as  $K\to \infty$. Both relations (\ref{VMD},\ref{fpisum1}) hint to the fact that a realistic model
of QCD will require at least a few sites to reproduce the data.

\section{Results}

The aim of this section is to quantify the agreement with the data of several continuous and discrete
models. All the results are summarized in table \ref{table} at the end of the paper.
The main outcome is that many different models, including flat space, have a good agreement with experimental
results. Indeed, as we will see,  only vague features of the metric are required to get fair predictions.

In order to compare different models the first question to address is how to quantify their
agreement with data. This is not obvious and involves a certain arbitrariness because the observables we
compute are subject to a theoretical error that we can only estimate through power counting
arguments. From this estimate we would expect typical relative errors in the $10$ to $30\%$
range, larger than the experimental error for many observables.
As in \cite{Erlich:2005qh,Pomarol:2008aa} we proceed as follows: We restrict
to the more precisely measured observables (namely, to those whose relative experimental error
is less than $10\%$) for which the experimental error can be neglected in comparison with the
theoretical one. We then define our error by the root mean square error estimate (RMSE).
Denoting with $\mathcal{O}^i_\text{exp}$ and $\mathcal{O}^i_\text{th}$ the experimental and theoretical values
of each observable, we have
\begin{equation}
\text{RMSE} = \sqrt{ \frac{1}{N_o} \sum_i \left(\frac{\mathcal{O}^i_\text{exp}-\mathcal{O}^i_\text{th}}{\mathcal{O}^i_\text{exp}}\right)^2 }\,,
\label{RMSE}
\end{equation}
where $N_o$ is the number of observables. The RMSE depends on the parameters of the model which
we fix to their best--fit value by minimization; the minimum value of the RMSE is our measure of the error and is
reported in the tables which follow for each different model.

The statistical meaning of this procedure is the following: assuming that the theoretical error of each
observable is $\Delta\mathcal{O}^i_\text{th}=\xi\mathcal{O}^i_\text{exp}$ we have
\begin{equation}
\text{RMSE} = \xi\, \sqrt{ \frac{1}{N_o} \sum_i \left(\frac{\mathcal{O}^i_\text{exp}-\mathcal{O}^i_\text{th}}{\Delta\mathcal{O}^i_\text{th}}\right)^2 } = \xi\,\sqrt{\frac {N_o-d} {N_o}}\, \chi_\text{red}\,,
\end{equation}
where $d$ is the number of fit parameters. For a good quality fit $\chi_\text{red}^2\sim 1$ so the RMSE ($d$ will be significantly smaller than $N_o$) determines the size of the corrections necessary for agreement with the data. The assignment of the common relative error $\xi$ is the main source
of ambiguity in the procedure; there is obviously no rigorous way to establish whether a $\xi$ error should be given to an observable $\mathcal{O}^i$ or, say, to its square ${\mathcal{O}^i}^2$. This translates in an ambiguity in the choice of the list of observables to be used in eq.~(\ref{RMSE}).
Our criterion was to use as observables the parameters of the corresponding matrix elements. We have checked that other choices, which of course produce slightly different results, leave the general conclusions unchanged especially for what concerns the comparison among different models.

For each model we computed a total of 18 observables, 13 of which fulfill the ``precision" criterion explained above and are used to compute the RMSE error of eq.~(\ref{RMSE}). The 5 additional observables (shown in table~\ref{table}) that we compute but do not use in computing the RMSE are: the $L_{1}$, $L_2$, $L_3$ chiral Lagrangian parameters, which are not well enough measured, the $\rho'$ mass and the $a_1$ decay constant. We expect the $\rho'$ (being it heavy) to be subject to larger theoretical errors than the other observables and this is why we  exclude its mass from the RMSE. Concerning the $F_{a_1}$, we do not use the usually quoted experimental value of $144$~MeV \cite{Isgur:1988vm} because this value is extracted from the $\tau$ decay in a model which is different from ours. Indeed, by computing the $\tau\to 3\pi\nu$ rate in our case (in which the branching ratio of $a_1\to 3\pi$ is one and no other resonances contribute to the process) we find a slightly bigger value, around $165$ MeV. Moreover, lattice calculations give a central value of $F_{a_1}=170$ MeV \cite{Wingate:1995hy}. All this suggests that $F_{a_1}$ is not known precisely enough to be included in the RMSE.

\subsection{5d models}

In the context of 5d geometries we present 4 representative models: flat space, AdS, Sakai-Sugimoto and for comparison
a decreasing warp factor less steep than AdS, $a(z)=\sqrt{L/z}$. For a given geometry the model depends only on two free parameters: $g_5$ and $L$ (as explained in section 2 we fix the gauge coupling of $U(1)$ and $SU(2)$ to be equal).
Even though for flat and AdS space analytic formul\ae{} can be used, in all the cases we compute the observables by first discretizing the model (through the dictionary (\ref{dl}) as discussed in section 2.2) and solving numerically.
This method turns out to be very efficient numerically because the problem reduces to
the diagonalization of a $K\times K$ tri-diagonal matrix and allows to solve easily essentially any metric. Even for $K=500$ (much beyond the required accuracy) the full fit can be performed in a few seconds.

We now present the result for the most relevant cases:

\subsubsection{Flat Space}

The result of the global fit is reported in the table below. The main source of error of the model
arises from the flat space relation $m_{a1}=2 m_{\rho}$. This effect is also enhanced by using square masses.

\begin{table}[!ht]
\begin{center}
\[
\begin{array}{|l|c|c|c|c|c|c|c|c|c|c|c|c|c|}\hline
 & m_{\rho }{}^2 & m_{\omega }{}^2 & m_{a_1}{}^2 & f_{\pi } & F_{\rho } & F_{\omega } & g_{\rho \pi \pi } & g_{\rho \pi \gamma } & g_{\omega \pi \gamma } & g_{\omega \rho \pi } & 10^3 L_9 & 10^3 L_{10} & r_{\pi }{}^2 \\\hline
 \text{Exp.} & .78^2 & .78^2 & 1.2^2 & .087 & .15 & .14 & 6.0 & .22 & .72 & 15 & 6.9 & -5.5 & 12 \\
 \text{Th.} & .68^2 & .68^2 & 1.4^2 & .081 & .12 & .12 & 4.8 & .23 & .68 & 13 & 5.8 & -5.8 & 11 \\
 \text{Dev.} & -23 & -23 & 30 & -6 & -23 & -18 & -20 & 3 & -6 & -11 & -15 & 6 & -12 \\\hline
\end{array}
\]
\caption{Flat space. The global error is $17\%$. All the dimensionfull quantities are in powers of \small{GeV} and the deviations are in percent. Fitted parameters: $L^{-1} \approx 430$ MeV and $g_5^{-2} \approx 7.6$ MeV.}
\label{flat}
\end{center}
\end{table}
\subsubsection{AdS Space}

The global fit is,
\begin{table}[!ht]
\begin{center}
\[
\begin{array}{|l|c|c|c|c|c|c|c|c|c|c|c|c|c|}\hline
 & m_{\rho }{}^2 & m_{\omega }{}^2 & m_{a_1}{}^2 & f_{\pi } & F_{\rho } & F_{\omega } & g_{\rho \pi \pi } & g_{\rho \pi \gamma } & g_{\omega \pi \gamma } & g_{\omega \rho \pi } & 10^3 L_9 & 10^3 L_{10} & r_{\pi }{}^2 \\\hline
 \text{Exp.} & .78^2 & .78^2 & 1.2^2 & .087 & .15 & .14 & 6.0 & .22 & .72 & 15 & 6.9 & -5.5 & 12 \\
 \text{Th.} & .76^2 & .76^2 & 1.2^2 & .082 & .16 & .16 & 5.3 & .25 & .74 & 15 & 6.3 & -6.3 & 11 \\
 \text{Dev.} & -4 & -4 & 3 & -5 & 6 & 13 & -11 & 13 & 3 & 3 & -9 & 15 & -7 \\\hline
\end{array}
\]
\caption{AdS space. The global error is $8.5\%$. All the dimensionfull quantities are in powers of \small{GeV} and the deviations are in percent. Fitted parameters: $L^{-1} \approx 320$ MeV and $g_5^{-2} \approx 5.3$ MeV.}
\label{ads}
\end{center}
\end{table}

We see that the fit of AdS provides a better description of QCD data (particularly because of the right mass relations) even though flat space could also be considered satisfactory at the level of precision we are working. However the superiority of AdS should not be overestimated. First of all our absolute measure of the agreement is subject to a certain arbitrariness. Our estimate of the error assumes that all the theorical errors are equal, and also depends of the choice of observables (for example square masses rather than mass). Moreover as we discuss in section \ref{cutoffsection}, AdS requires a slighly lower cut-off than flat space as indicated by the value of $g_5$ in the fit. Having said this, the data do show a preference for a decreasing warp factor and AdS is already very close to the intrinsic error of the geometrical models (see below). In fact we showed by sampling a large number of geometries that the error is always larger than 7-8\%. More in general about any gently globally decreasing metric will lead to an error less than $20\%$. For example $a(z)=\sqrt{L/z}$ gives a fit as good as AdS (see the summary table \ref{table}).

\subsubsection{Sakai-Sugimoto Model}

For the case of the Sakai-Sugimoto model whose metric is given by \eqref{ssmetric} we find,

\begin{table}[!ht]
\begin{center}
\[
\begin{array}{|l|c|c|c|c|c|c|c|c|c|c|c|c|c|}\hline
 & m_{\rho }{}^2 & m_{\omega }{}^2 & m_{a_1}{}^2 & f_{\pi } & F_{\rho } & F_{\omega } & g_{\rho \pi \pi } & g_{\rho \pi \gamma } & g_{\omega \pi \gamma } & g_{\omega \rho \pi } & 10^3 L_9 & 10^3 L_{10} & r_{\pi }{}^2 \\\hline
 \text{Exp.} & .78^2 & .78^2 & 1.2^2 & .087 & .15 & .14 & 6.0 & .22 & .72 & 15 & 6.9 & -5.5 & 12 \\
 \text{Th.} & .78^2 & .78^2 & 1.2^2 & .080 & .18 & .18 & 5.5 & .26 & .79 & 17 & 6.6 & -6.6 & 12 \\
 \text{Dev.} & -1 & -1 & -2 & -8 & 22 & 31 & -8 & 20 & 10 & 12 & -4 & 20 & 2 \\\hline
\end{array}
\]
\caption{The Sakai-Sugimoto model. The global error is $14\%$. All the dimensionfull quantities are in powers of \small{GeV} and the deviations are in percent. Fitted parameters: $L^{-1} \approx 260$ MeV and $g_5^{-2} \approx 5.4$ MeV.}
\label{s-s}
\end{center}
\end{table}

It is interesting to see that the error is similar to flat space. The main discrepancy is in this case the decay constant of the $\omega$ (here taking a different coupling constant for the $U(1)$ would help to slightly fix this problem but would not improve the global error drastically). With respect to flat space this model reproduces more precisely the masses.

\subsection{Few sites}

We now turn to purely 4 dimensional models. In comparison to the model proposed by Georgi
in the context of the vector limit \cite{Georgi:1989xy} we show that the addition of a few sites
improves the agreement with QCD substantially.

As we discuss in the next section, to reproduce QCD the five-dimensional theories
considered have necessarily a low cut-off so that only a few resonances lie below the cut-off of the theory.
Since our effective description describes reliably only the first 3 or so Kaluza-Klein modes, it makes sense to
directly use a four dimensional model with few sites where only modes below the cut-off are included.
The difference between this and an appropriate 5d model amounts to higher dimensional operators
suppressed by the cut-off scale. For $K$ sites (\emph{i.e.} $K$ resonances) the model has $K+1$ parameters,
the local $\sigma$-model decay constants and the gauge couplings (as in the continuous case the parameters
of the $U(1)$ and $SU(2)$ sector are taken to be identical).

\subsubsection{2-Sites}

Fitting the 3 parameters of the model we get the table below. The 19\% error of this model is higher than the typical geometric model with decreasing metric. Note the poor predictions of $L_9$ and of the pion radius.

\begin{table}[!ht]
\begin{center}
\[
\begin{array}{|l|c|c|c|c|c|c|c|c|c|c|c|c|c|}\hline
 & m_{\rho }{}^2 & m_{\omega }{}^2 & m_{a_1}{}^2 & f_{\pi } & F_{\rho } & F_{\omega } & g_{\rho \pi \pi } & g_{\rho \pi \gamma } & g_{\omega \pi \gamma } & g_{\omega \rho \pi } & 10^3 L_9 & 10^3 L_{10} & r_{\pi }{}^2 \\\hline
 \text{Exp.} & .78^2 & .78^2 & 1.2^2 & .087 & .15 & .14 & 6.0 & .22 & .72 & 15 & 6.9 & -5.5 & 12 \\
 \text{Th.} & .77^2 & .77^2 & 1.1^2 & .085 & .12 & .12 & 4.7 & .21 & .64 & 11 & 4.7 & -4.7 & 8 \\
 \text{Dev.} & -3 & -3 & -23 & -3 & -17 & -12 & -21 & -2 & -10 & -25 & -32 & -15 & -35 \\\hline
\end{array}
\]
\caption{Discrete model with K=2. The global error is $19\%$. All the dimensionfull quantities are in powers of \small{GeV} and the deviations are in percent. Fitted parameters: $f_1\approx180$ MeV, $f_2\approx120$ MeV and $g_1\approx8.8$.}
\label{K2}
\end{center}
\end{table}

\subsubsection{3-Sites}

In this case the number of free parameters is 4. The best fit corresponds to a 12\% error.
\begin{table}[!ht]
\begin{center}
\[
\begin{array}{|l|c|c|c|c|c|c|c|c|c|c|c|c|c|}\hline
 & m_{\rho }{}^2 & m_{\omega }{}^2 & m_{a_1}{}^2 & f_{\pi } & F_{\rho } & F_{\omega } & g_{\rho \pi \pi } & g_{\rho \pi \gamma } & g_{\omega \pi \gamma } & g_{\omega \rho \pi } & 10^3 L_9 & 10^3 L_{10} & r_{\pi }{}^2 \\\hline
 \text{Exp.} & .78^2 & .78^2 & 1.2^2 & .087 & .15 & .14 & 6.0 & .22 & .72 & 15 & 6.9 & -5.5 & 12 \\
 \text{Th.} & .77^2 & .77^2 & 1.1^2 & .082 & .14 & .14 & 5.0 & .23 & .70 & 13 & 5.3 & -5.3 & 9 \\
 \text{Dev.} & -3 & -3 & -14 & -5 & -4 & 3 & -17 & 6 & -3 & -15 & -23 & -3 & -21 \\\hline
\end{array}
\]
\caption{Discrete model with K=3. The global error is $12\%$. All the dimensionfull quantities are in powers of \small{GeV} and the deviations are in percent. Fitted parameters: $f_1\approx270$ MeV, $f_2\approx130$ MeV, $g_1\approx7.3$ and $g_2\approx10$.}
\label{K3}
\end{center}
\end{table}

Therefore a model with 3-sites and nearest neighbor interactions provides an excellent fit
of the QCD observables considered. For comparison, had we performed an
analog fit in the Georgi's model \cite{Georgi:1989xy} (clearly removing certain observables such as the ones associated to the $a_1$) the error of the fit would have been more than 30\%. Note that the best fit drives the decay constants
in a region where they are hierarchical. This somewhat resembles the case of the AdS metric.
In fact (see next section) the low value of the decay constant at the infrared worsens the calculability of the theory, but on the other hand a non generic choice of parameters is required to improve the fit, so there seems to be a balance between calculability and low error of the model.

We also considered the case of 4 sites. In this case we find an error less than 10\%, as low as the one for AdS (see the general table (\ref{table})). We should note however that such agreement emerges at the price of rather large values for some of the $g$'s and again hierarchical values of  the $f$'s. Increasing further the number of sites does not make the error decrease much further so we conclude that low energy QCD can be well described by a discrete model with 3-4 sites.

\subsection{The intrinsic error}

All the models that we have considered share common predictions, like for instance
the relations between the $L_i$ coefficients of the $\mathcal{O}(p^4)$ chiral Lagrangian, the sum rules
discussed in sect.~2.3.3 and the equality of the $\rho$ and $\omega$ masses. These predictions are
phenomenologically quite successfull and partially explain why different models show a
comparable and fair agreement with the data. Our model--independent relations, however, are not
perfect and therefore produce an irreducible intrinsic error which is impossible to eliminate,
even allowing for completely arbitrary geometries. We quantify this intrinsic error as 7-8\%,
which is the minimum error we obtain by scanning the parameter space of the discrete model
with many ($K=100$) sites. The intrinsic error is not far from the one of AdS or
$\sqrt{L/z}$, meaning that these geometries have the non--trivial property of practically
minimizing the error.

It is natural to ask what fraction of the total error has to be
really ascribed to the choice of the geometry and what part is due to the
unavoidable intrinsic error. To answer this question we consider a restricted set
of observables for which the tension due to the model--independent predictions is eliminated.
These are the masses of the $\rho$ and of the $a_1$, the decay
constants of pion, $\rho$ and $\omega$ and $g_{\omega\rho\pi}$. The fit gives now
an error of $4.8\%$ and $6.5\%$ for AdS and $1/\sqrt{z}$
respectively, while for the Sakai-Sugimoto model the error practically does not change
and for flat space it even increases by about 1\% due to the bad mass relations.
The scan of the discrete model parameter space gives now a minimum error which
is essentially zero, showing that indeed we have completely eliminated the intrinsic
 error with this new choice of the observables.
These results have been obtained by using more precise  experimental values
and allowing for different couplings in the $U(1)$ and $SU(2)$ sectors, {\it{i.e.}} by
fitting the parameter $\alpha$ of eq.~(\ref{action}) instead of fixing it to one.
At this level of precision small deviations of $\alpha$ from one have indeed a
significant impact.

Some of the relations which produce the intrinsic error are obvious:
the mass of the $\omega$ is the same as the mass of the $\rho$,
 $L_9$ and $L_{10}$ are fixed by $r_\pi$ and $f_\pi$ and the couplings $g_{\rho\pi\gamma}$ and
$g_{\omega\pi\gamma}$ are expressed in terms of $g_{\rho\pi\pi}$ and $f_\pi$ (see sect.~2).
 Another relation, already discussed below eq.~(\ref{fpisum1}), is that
\begin{equation}
3\;\frac{{f_\pi}^2\, {g_{\rho\pi\pi}}^2}{{m_\rho}^2} < 1\,,
\label{b1}
\end{equation}
independently on the warp factor, while the experimental value is about $1.3$.
Also, the sum in eq.~(\ref{pionradius}) which gives the radius of the pion is essentially
always saturated by the first resonance. We therefore find the relation
\begin{equation}
6\, \frac{ F_\rho\; g_{\rho\pi\pi}}{{m_\rho}^3 {r_\pi}^2} = 1\,,
\end{equation}
but the experimental value is so close ($1.02$) that even the small deviations due to the rest of the
sum in eq.~(\ref{pionradius}) allow the experimental result to be exactly reproduced.

It is important to remark that the model--independent predictions discussed in this
section, which lead to the minimal intrinsic error, are only valid at the tree--level
and will be modified by higher order corrections. A non--zero value of
$L_9+L_{10}$, for instance, will be generated by loop corrections and its size
is expected, by simple power counting arguments, to be consistent with the
experimental one. The vanishing of $L_9+L_{10}$ at tree--level  is related with the tree--level
absence of the $a_1\rightarrow\pi\gamma$ decay, which is another feature of
our model. This relation is lost, however, at the loop level and indeed even in the
HLS case, in which the $a_1$ meson is not present, a non--vanishing $L_9+L_{10}$
is generated \cite{Harada:2003jx}. It would be interesting to etablish
whether radiative corrections induce a realistic $a_1\rightarrow\pi\gamma$
decay in the models we have considered.

\section{Calculability}
\label{cutoffsection}

One important feature of extra-dimensional descriptions of QCD is
provided by the calculability of these models.
We now wish to consider this aspect in detail. The arguments below
become good estimates in the case of a large number of colors.
We consider the case of a flat geometry where the power counting is
transparent. Naively, neglecting boundary effects, the five
dimensional theory has a regime of validity up
to the scale $\Lambda\approx 16 \pi^2 /g_5^2$. Given the relations,
\begin{eqnarray}
f_\pi^2&=& \frac 2 {g_5^2 L} \nonumber \\
m_\rho&=&\frac {\pi}{ 2 L}
\end{eqnarray}
and the fact that $4 \pi f_\pi\approx\sqrt{N_c} m_{\rho}$ it follows
that $1/g_5^2$ is linear in $N_c$.
In absence of other interactions this would imply that the number of
weakly coupled resonances below the cut-off scale is also roughly $N_c$.

Due to the presence of the Chern-Simons term, necessary to reproduce
the anomalies of the theory, the number of weakly coupled resonances
is however smaller for $N_c$ large, see also \cite{Pomarol:2008aa}. This is due to
the fact that the coefficient of the Chern-Simons
term scales with $N_c$. This interaction is more relevant than the
pure non-abelian gauge interaction
at large $N_c$ and lowers the maximum possible cut-off of the theory.
\footnote{Independently of anomalies in the gravitational dual of
theories such as $N=4$ Super-Yang-Mills one finds
$M_5/m_s \propto N_c^{\frac 2 3}$ where $m_s$ is the string scale. It
is therefore reasonable to expect in general that this
cutoff corresponds to a physical scale.} Considering the growth with energy of scattering
amplitudes of gauge bosons in 5D the new cut-off can be estimated to be
\begin{equation}
\Lambda_{MAX} =   \left(\frac \kappa {N_c}\right)^{\frac 2 3} \frac {16\pi^2} {g_5^2}\,,
\end{equation}
where $\kappa$ is a numerical factor of the order few for the most relevant interaction 
(see eq.~(\ref{action}). This means that the maximum number of weakly coupled resonances 
that the higher dimensional theory can possibly describe scales as
\begin{equation}
N_{max}\propto N_c^{1/3}\,.
\label{kkmax}
\end{equation}
What is important for the calculability of the 5d theory is that
$N_{max}$ still grows, allowing to have a weakly coupled deformation of the theory by taking
$N_c$ large. Despite the fact that the Chern-Simons lowers the cut-off,
the couplings of the mesons respect the correct scaling of large $N_c$ QCD.
This follows from the fact that since $1/g_5^2$
is proportional to $N_c$ we can factor out the number of colors in the
five-dimensional action. As a consequence interactions of the KK modes
are suppressed at large $N_c$ and the appropriate countings are reproduced.
Using the previous rules for the case of $N_c=3$ considered in this paper
we note that the Chern-Simons does not lower cut-off in this case and 3 resonances
are weakly coupled in our effective description.

It is interesting to see how the above results can be understood in
the deconstructed version of the
theory. The basic equations in this case are given by
\begin{eqnarray}
f_\pi^2&=& \frac {f^2}{K}\,,\nonumber \\
m_\rho&=& \frac {g f}{K}\,,
\end{eqnarray}
where $K$ is the number of sites. It follows that $16 \pi^2 K \approx
g^2 N_c$.
Since $g<4 \pi$ for the gauge theory to make sense, this already tells
us that the 4d
description cannot contain more than $\mathcal{O}(N_c)$ states that
are weakly coupled.
However, as in the 5d case, this upper bound cannot be saturated due
to the interactions associated
with the anomalies. The cut-off can be computed locally by looking at
the scale where the individual $\sigma-$models
become strongly coupled. In the discretized version the Chern-Simons
term decomposes into gauged WZW terms
with coefficients $N_c$ at each site (see appendix). The most relevant
interaction arises from the term
\begin{equation}
\frac {N_c\, g^2} {16 \pi^2}  \frac {\pi \,F \wedge F}{ k\,f}\,,
\end{equation}
which becomes strongly coupled at the scale $\Lambda \approx 64  \pi^3 k\,f/
(N_c g^2)$. Requiring that the heaviest state
lies below the cut-off one finds that the number of weakly coupled
resonances is maximized by taking
\begin{equation}
g\approx \frac {4\pi}{N_c^{\frac 1 3}}\,.
\end{equation}
This implies, as in the continuous version, the relation (\ref{kkmax}).
Note that contrary to the pure gauge case
reproducing the 5d cut-off demands that the local gauge coupling goes
to zero.
Therefore the geometrical picture and the deconstructed version with
nearest neighbor interactions and a number of sites
$\approx N_c^{\frac 1 3}$  are indeed equivalent as effective field
theories.

Similar considerations can be repeated in the case of a generic
geometry/deconstructed model.
An interesting fact is that, working with $m_\rho$ and $f_\pi$ fixed,
flat space actually maximizes the cut-off defined by
the 5d gauge coupling. For example to reproduce the experimental values of
$f_\pi$ and $m_{\rho}$ AdS requires $g_5^2$ which is 30\%
larger than flat space. In fact the result of our fit is strongly
correlated to the dependence of the formal cut-off of the theory.
Of course such statements should be taken with great care as there is
no parametric separation between the
cut-off in flat and AdS space. However for the case of interest,
$N_c=3$, where the calculabilty of the model is
inevitably limited, such a numerical difference has an impact.

\section{Discussion}

In this paper we compared several different holographic models of QCD to quantify their
agreement with data. It is rather remarkable that, compatibly with the corrections expected in these type of models,
essentially any smooth decreasing metric and even flat space provide a good fit of the low energy mesonic observables.
This suggests that there is no special virtue in the choice of the AdS metric originally inspired by
the AdS/CFT correspondence. Nevertheless it is interesting to note that, in the class of
models we considered, AdS almost minimizes the error while flat space and
Sakai-Sugimoto models give a similar error which is higher than AdS.

Perhaps the most interesting outcome of our analysis is that the successful results of the
holographic models of QCD are already reproduced by a four dimensional model with $\cal{O}$(3) sites.
This is to be expected because all the five-dimensional models
are characterized by a low cut-off fixed by the experimental values of $m_{\rho}$ and $f_\pi$.
As a consequence only $\cal{O}$(3) resonances can be reliably described within the effective theory
and a macroscopic notion of geometry is lost. In this sense it might be useful to consider
holographic theories of QCD as discrete theories containing only the modes lying below
the cut-off of the theory. These are generalizations of the model proposed by Georgi in \cite{Georgi:1989xy}.
What is crucial phenomenologically is the  structure of nearest neighbor interactions of the discrete
theory/extra-dimension which also ensures an enhanced calculability of the theory.

Among possible extensions of our analysis the most important would be the inclusion of the
baryon sector of the theory. As shown in \cite{Pomarol:2008aa} the baryons arise in
the 5d picture as topological solitons analogous to 4d skyrmions. The extra-dimension
allows again to make such configurations calculable contrary to their 4d cousins. The same will be true
for any metric and discrete model considered here, even though explicit solutions
might not be easy to find. As in \cite{Pomarol:2008aa}, it would be interesting to consider
the impact of the baryon observables on the global fit.

We do not expect one model to be strongly preferred. Due to the small value of $N_c$ in QCD the calculability of the theory is always limited so many different models will always give comparable predictions. In this regard the AdS/QCD like models
considered in this paper make sharp predictions for a world with a large number of colors already at tree-level,
so if such data were available from lattice computations, it would allow to test whether such approach has some truth or should be rejected, and which geometry is favored. Alternatively one could try to compute next-to-leading order corrections, paying the price of adding new independent parameters, to check whether the intrinsic error of these models is reduced and if a preference for some geometry emerges.

\vspace{0.5cm}
{\bf Acknowledgments}

We would like to thank G.~Isidori and R.~Rattazzi for interesting conversations.
Special thanks to Giuliano Panico and to Alex Pomarol for many discussion and
comments on the material contained in this paper. We would also like to thank
Gregoire Gallet for collaboration at the early stages of this work.

\vspace{-.3cm}

\appendix

\section{Anomalies and the Wess-Zumino-Witten term}

In this appendix we show how the QCD anomalies can be incorporated in the discrete model by the
introduction of a gauged WZW at each link.
Due to anomalies the effective action of large--$N_c$ QCD must transform as
\beq
\delta S\,=\,\frac{N_c}{24\pi^2}\int\left[{\overline\omega}_{4}^1({\boldsymbol\alpha}_L,{\mathbf l})
-{\overline\omega}_{4}^1({\boldsymbol\alpha}_R,{\bf r})
\right] \,,
\label{an}
\eeq
where the 4--form ${\overline\omega}_{4}^1({\boldsymbol\alpha},{\bf A})$,
whose exact definition is irrelevant for the moment,
represents the QCD anomaly and will be given in eq.~(\ref{ombar}).
Infinitesimal local chiral symmetry transformations, which correspond
to ${\bf L}={\bf L}_1$ and ${\bf R}={\bf R}_{K+1}$ in eq.~(\ref{ggroup}),
are parametrized by ${\boldsymbol\alpha}_{L,R}$ as
${\bf L}\simeq\I+i{\boldsymbol\alpha}_{L}$, and similary for ${\bf R}$.
Notice that in writing eq.~(\ref{an}) we have ignored the $U(1)_A$--$SU(N_c)_c$--$SU(N_c)_c$
anomaly, which is the one responsible for the $\eta'$ mass.
This is justified in the large--$N_c$ limit where the latter anomaly is subleading
\cite{Witten:1979vv} and the $\eta'$ is massless as it will be in our model.
Needless to say, the anomaly
is only defined up to the variation of local counterterms, so that we had to make
a choice to write eq.~(\ref{an}). With our choice the anomaly is entirely in the
$U(1)$ part (\emph{i.e.}, it vanishes for $SU(2)$ transformations, see eq.~(\ref{ombar}))
and assumes
the ``Left minus Right'' form. Having made a choice does not imply, however, a loss
of generality:
any other form of the anomaly
could be obtained by adding local counterterms to the Lagrangian we will derive
assuming eq.~(\ref{an}). \footnote{If we had, for instance, to put the anomaly in the
standard form in which the vector subgroup is anomaly--free we would just
 add the Bardeen counterterm of eq.~(\ref{bardeen}).}

According to the previous discussion, an ``anomalous'' non gauge--invariant term,
 with variation given by eq.~(\ref{an}), has to be added to the action;
we will denote the latter as $S_{A}$.
In order to construct $S_{A}$ let us assume that each of the
$K+1$ $\sigma$--models is anomalous, exactly as it would be if it
represented the effective description of a large--$N_c$ QCD--like
theory. If this is the case the action, prior to the gauging,
contains a gauged WZW term for each link
\beq
\sum_{k=1}^{K+1}\Gamma_{WZW}\left[{\bf\Sigma}^k,{\bf l}^k,{\bf r}^k\right]\,,
\label{wzwk}
\eeq
where $\Gamma_{WZW}$ will be defined in eq.~\eqref{wzw1}. The WZW terms,
according to eq.~(\ref{anst}), produce an anomaly
\beq
\sum_{k=1}^{K+1}\delta\Gamma_{WZW}\left[{\bf\Sigma}^k,{\bf l}^k,{\bf r}^k\right]
\,=\,\frac{N_c}{24\pi^2}\int\sum_{k=1}^{K+1}\left[{\overline\omega}_{4}^1({\boldsymbol\alpha}_L^k,{\mathbf l}^k)
-{\overline\omega}_{4}^1({\boldsymbol\alpha}_R^k,{\bf r}^k)
\right] \,.
\label{ank}
\eeq
Notice that the variation (\ref{ank}) vanishes for a global transformation and eq.~(\ref{wzwk})
is still invariant under the entire global group of eq.~(\ref{ggroup}).

We must now, as we did before with the Lagrangian (\ref{lag_ung}), gauge the vector
combination ${\bf A}^k$ of the ${\bf l}^{k+1}$ and ${\bf r}^{k}$ sources.
To this end we replace ${\bf l}^{k}$ with ${\bf A}^{k-1}$ and ${\bf r}^{k}$
with ${\bf A}^{k}$ in the action (\ref{wzwk}) and obtain
\beq
S_{A}\,=\,\sum_{k=1}^{K+1}\Gamma_{WZW}\left[{\bf\Sigma}^k,{\bf A}^{k-1},{\bf A}^k\right]\,.
\label{scsd}
\eeq
Under the local groups that we would like to gauge and under the ungauged
``end--points'' groups which correspond to the chiral QCD ones the action
$S_{A}$ transforms as
\beq
\delta S_{A}\,=\,\frac{N_c}{24\pi^2}\int\sum_{k=1}^{K+1}\left[{\overline\omega}_{4}^1({\boldsymbol\alpha}^{k-1},{\mathbf A}^{k-1})
-{\overline\omega}_{4}^1({\boldsymbol\alpha}^k,{\bf A}^k)
\right]\,=\,\frac{N_c}{24\pi^2}\int\left[{\overline\omega}_{4}^1({\boldsymbol\alpha}_L,{\mathbf l})
-{\overline\omega}_{4}^1({\boldsymbol\alpha}_R,{\bf r}) \right]\,.
\eeq
The second equality follows from the identifications ${\boldsymbol\alpha}^0={\boldsymbol\alpha}_L$,
${\boldsymbol\alpha}^{K+1}={\boldsymbol\alpha}_R$, $\A^0={\bf l}$ and $\A^{K+1}={\bf r}$
and from the fact that the anomaly of each group, except the first and the
last ones, receives equal and opposite contributions from the links to its left and to its right.
This ensures that the anomaly is cancelled for the vector groups, which can therefore be
safely gauged, and that $S_{A}$ has the correct anomaly required by eq.~(\ref{an}).
Moreover, thanks to our construction, $S_{A}$ is in the nearest--neighbor form of eq.~(\ref{nnl})
like the rest of the action. Notice that for the cancellation of the vector anomaly it has been
crucial that we chose the  anomalies of all the $\sigma$--models to be equal (\emph{i.e.},
to correspond to the same number of colors $N_c$) and each of them to be in the
``Left minus Right'' form. It is easy to realize that if we had started from any other form
of the anomaly, as long as it was of the ``Left minus Right'' type, we would have obtained the
same $S_{A}$ action.

We now need, in order to write the explicit form of $S_A$ of eq.~(\ref{cs}), to work out the
$\Gamma_{WZW}$ action. This is the subject of the next subsection.

\subsubsection*{Gauged WZW for $\mathbf{U(2)}$}

The general expression for the gauged WZW term in the case of $N_f$ flavors, \emph{i.e.} of $U(N_f)$ or $SU(N_f)$ chiral
groups, is well--known and can be found in \cite{Witten:1983tw}.
In the two flavors case that we are considering, however, the
general formula can be strongly simplified and, up to gauge--invariant operators, the WZW assumes a much
simpler form. It is more convenient and more interesting, instead of starting from the general formula
and simplifying it, to derive from scratch the WZW term for two--flavor QCD and this is what we will do in
the following.
We will use a form notation in which $A \equiv -\ii A_\mu \dd x^\mu$, where $A$ denotes here any Left or Right gauge
field, and the field strength is a 2-form $F = \dd A + A^2$.

Our starting point is the chiral QCD anomaly which, by a standard text--book calculation can be written as
\beq
\delta \Gamma\left[{\mathbf l},{\mathbf r}\right]\,=\,\frac{N_c}{24\pi^2}\int\left[{\omega}_{4}^1({\boldsymbol\alpha}_L,{\mathbf l})
-{\omega}_{4}^1({\boldsymbol\alpha}_R,{\bf r})
\right] \,,
\label{anst}
\eeq
where $\delta\Gamma$ denotes the variation under local chiral transformations ${\boldsymbol\alpha}_{L,R}$ of the
 effective action for the $U(2)_L\times U(2)_R$ sources ${\mathbf l}$ and ${\mathbf r}$. The
${\omega}_{4}^1$ 4--form is defined as
\beq
\displaystyle
{\omega}_{4}^1({\boldsymbol\alpha},{\mathbf A})\,=\,\tr{{\boldsymbol\alpha}\,\dd\left(\A\dd \A+\frac12 \A^3\right)}\,,
\eeq
and eq.~(\ref{anst}) provides the standard ``symmetric'' form of the anomaly. It is customary to define the
Chern--Simons 5--form $\omega_5$ by regarding the ordinary 4d space--time as the boundary of a fictitious 5d
and extending the sources to 5d gauge fields in this space. It is
\beq
\displaystyle
{\omega}_5({\mathbf A})\,=\,\ii\,\tr{\A\,\left(\dd \A\right)^2\,+\,\frac32 \A^3\dd\A\,+\,\frac35\A^5}\,,
\eeq
and the variation of $\omega_5$ is the exterior derivative of the anomaly: $\delta_{\boldsymbol\alpha}\omega_5=\dd{\omega}_{4}^1$.

All the equations above hold in general, for any number of flavors, but an important simplification occurs
if we specialize to the case of $N_f=2$ in which $\A=\I/2{\widehat{A}}+A$ is an $U(2)$ gauge field. The CS
becomes
\begin{equation}
 \omega_5({\mathbf A}) \,=\, \ii\,\frac{1}{4}\, \hat{A} (\dd \widehat{A})^2 + \ii\frac{3}{2}\, \widehat{A} \tr{F^2}
\,+\,\ii\,\dd\left[\widehat{A}\tr{A\,F-\frac14A^3}\right]\,\equiv\,{\overline{\omega}}_5({\mathbf A})\,+\,\dd X
\,.
\end{equation}
This means that, by adding the local counterterm $X$ to the effective action ({\it{i.e}}, by a change of the regulator),
the anomaly (\ref{anst})) can be put in the form
\beq
\delta \Gamma\left[{\mathbf l},{\mathbf r}\right]\,=\,\frac{N_c}{24\pi^2}\int\left[{\overline\omega}_{4}^1({\boldsymbol\alpha}_L,{\mathbf l})
-{\overline\omega}_{4}^1({\boldsymbol\alpha}_R,{\bf r})
\right] \,,
\label{annew}
\eeq
where the 4--form
\beq
\displaystyle
{\overline\omega}_{4}^1({\boldsymbol\alpha},{\bf A})\,=\,
\frac14{\widehat{\alpha}}\left(\dd{\widehat{A}}\right)^2\,+\,
\frac32{\widehat{\alpha}}\tr{F^2}\,,
\label{ombar}
\eeq
is defined, up to an irrelevant total differential which cancels under integration, from the relation
\beq
\delta_{\boldsymbol\alpha}{\overline\omega}_5=\dd{\overline\omega}_{4}^1\,.
\eeq
In the alternative form (\ref{ombar}), the anomaly is entirely in the $U(1)$ part and the effective action
$\Gamma$ is $SU(2)_L\times SU(2)_R$--invariant.

In the low--energy effective action for the Goldstone bosons $U(2)$ matrix ${\bf\Sigma}$ the anomaly (\ref{annew})
is reproduced by a suitably designed gauged WZW term, $\Gamma_{WZW}\left[{\bf\Sigma},{\bf l},{\bf r}\right]$, which is
such that its variation respects eq.~(\ref{ombar}). Following ref.~\cite{Chu:1996fr},
a term with this property is
\begin{equation}
\displaystyle
 \Gamma_{WZW}\left[{\bf\Sigma},{\bf l},{\bf r}\right] = \frac{N_c}{24\pi^2} \int \left\{
\left[
{\overline \omega}_5({\bf l}) - {\overline \omega}_5({\bf r}) \right] \,-\,
\left[ {\overline \omega}_5({\bf l}^{(\hc{\bf\Sigma})})-{\overline \omega}_5({\bf r})
+ \dd B_4({\bf l}^{(\hc{\bf\Sigma})},{\bf r}) \right] \right\}\,,
\label{wzw0}
\end{equation}
where ${\bf l}^{(\hc{\bf\Sigma})}$ denotes the Left source gauge--rotated by the inverse
Goldstone matrix $\hc{\bf\Sigma}$. Given that ${\bf\Sigma}$ transforms as
${\bf\Sigma}\rightarrow g_L{\bf\Sigma}\hc{g_R}$, ${\bf l}^{(\hc{\bf\Sigma})}$
transforms in the same way as the ${\bf r}$ source does.

The way eq.~(\ref{wzw0}) works is
not difficult to explain: the first square bracket alone already gives the anomaly but is
not a closed form so that it depends on the unphysical extention of the sources (and of the
Goldstones) in the extra dimension. The second square bracket must be there to give a closed
form when summed with the first one, and must be gauge--invariant. It clearly accomplishes the
first task as $\dd{\overline{\omega}}_5=\dd\omega_5=\tr{{\bf F}^2}$ but for the second one we
have to choose a suitable ``Bardeen counterterm'' 4--form $B_4$. This must be such that
under vector transformations (remember that ${\bf l}^{(\hc{\bf\Sigma})}$ transforms as
${\bf r}$) ${\overline \omega}_5({\bf l})-{\overline \omega}_5({\bf r})
+ \dd B_4({\bf l},{\bf r})$ is invariant. This means that
\beq
\delta_VB_4({\bf l},{\bf r})\,=\,-\,\left[{\overline\omega}_{4}^1({\boldsymbol\alpha},{\bf l})
-{\overline\omega}_{4}^1({\boldsymbol\alpha},{\bf r})
\right]\,.
\eeq
Given that the anomaly is enterely in the $U(1)$ part, $B_4$ must be invariant under $SU(2)$
vector transformations and moreover, due to parity invariance, it must be odd under the
${\bf l}\rightarrow{\bf r}$ operation. Also, it must contain only one derivative as its
gauge variation must contain two. The only two independent terms with those feature
which contain $SU(2)$ fields and are not gauge invariant are
$\ii{\widehat v}\tr{a(L+R)}$ and $\ii{\widehat v}\tr{a^3}$ where
${\bf v}=({\bf l}+{\bf r})/2$ and ${\bf a}=({\bf l}-{\bf r})/2$. The coefficients are extracted
by direct calculation and the result is
\begin{equation}
 B_4({\bf l},{\bf r}) = -\frac{\ii}{4}\, \left( {\widehat{l}}\, {\widehat{r}}\,
( \dd{\widehat{l}} + {\dd\widehat{r}} ) \right) + \frac{\ii}{4}\, ({\widehat{l}}
+ {\widehat{r}}) \tr{(l - r) \left[ 3\, (L + R) - (l - r)^2 \right]}\,.
\label{bardeen}
\end{equation}

Applying now eq.~(\ref{wzw0}) and after some simple algebra we can write
\beq
\displaystyle
\Gamma_{WZW}\left[{\bf\Sigma},{\bf l},{\bf r}\right] \,&=&\,
 \ii \frac{N_c}{96\pi^2} \int \left[ {\widehat{l}}\, {\widehat{r}}\, ( \dd{\widehat{l}}
 + \dd{\widehat{r}} ) - \ii{\widehat{\Pi}} \left( (\dd {\widehat{l}})^2 + (\dd {\widehat{r}})^2 +
\dd {\widehat{l}}\, \dd {\widehat{r}} \right) - 3\ii\, {\widehat\Pi} \tr{L^2 + R^2}\right. \notag\\
&&\left.
 +\ ({\widehat{l}} + {\widehat{r}}) \tr{3 \Bigl( L \Sigma D\hc{\Sigma} - R \hc{\Sigma} D\Sigma \Bigr) + \Bigl( \hc{\Sigma} D\Sigma \Bigr)^3 } \right]\,,
\label{wzw1}
\eeq
where we have separated ${\bf\Sigma}={\widehat{\Sigma}}\Sigma$ in the $U(1)$ and $SU(2)$ component
and parametrized ${\widehat{\Sigma}}$ as ${\widehat{\Sigma}}=\exp{\ii{\widehat\Pi}}$.
Notice that, as known, no ungauged WZW term is present in the $U(2)$ case and, consequently,
$\Gamma_{WZW}$ in eq.~(\ref{wzw1}) is enterely written as an ordinary 4d integral.

The WZW term, in the usual NDA counting of the chiral Lagrangian in which gauge
fields and derivatives both count as ${\mathcal{O}}(p)$, is an
${\mathcal{O}}(p^4)$ term and its coefficient is perfectly consistent with the
NDA rule. \footnote{One has to remember that physical cut of the cutoff of the $\sigma$--model,
\emph{i.e.} the scale at which the $\rho$ meson enters, is
$\Lambda_{\chi SB}=4\pi f_{\pi}/\sqrt{N_c}$ at large--$N_c$, and not $4\pi f_\pi$.}
At the same order, several gauge--invariant operators appear, which are listed in general in
ref.~\cite{Kaiser:2000gs}
for the $U(3)$ chiral group. The WZW term, therefore, is in any sense unique given that we could have
added to eq.~(\ref{wzw1}) any combination of these ${\mathcal{O}}(p^4)$ gauge--invariant operators.

\newpage\begin{landscape}
\begin{table}[!ht]
\begin{center}
\small
\[
\begin{array}{|l|r|rr|rr|rr|rr|rr|rr|rr|}\hline
 &&\multicolumn{2}{c|}{\text{Flat}}&\multicolumn{2}{c|}{\text{AdS}}&\multicolumn{2}{c|}{\text{Sak.-Sug.}}&\multicolumn{2}{c|}{\sqrt{L/z}}&\multicolumn{2}{c|}{\text{K=2}}&\multicolumn{2}{c|}{\text{K=3}}&\multicolumn{2}{c|}{\text{K=4}} \\
 \text{Observable} & \text{Exp.} & \text{Th.} & \text{Dev.} & \text{Th.} & \text{Dev.} & \text{Th.} & \text{Dev.} & \text{Th.} & \text{Dev.} & \text{Th.} & \text{Dev.} & \text{Th.} & \text{Dev.} & \text{Th.} & \text{Dev.} \\ \hline\hline

 m_{\rho }{}^2 & .78^2 & .68^2 & -23 & .76^2 & -4
& .78^2 & -1 & .74^2 & -11
& .77^2 & -3 & .77^2 & -3 & .76^2 & -5 \\

 m_{\omega }{}^2 & .78^2 & .68^2 & -23 & .76^2 & -4
& .78^2 & -1 & .74^2 & -11
& .77^2 & -3 & .77^2 & -3 & .76^2 & -5 \\

 m_{a_1}{}^2 & 1.2^2 & 1.4^2 & 30 & 1.2^2 & 3
& 1.2^2 & -2 & 1.3^2 & 14
& 1.1^2 & -23 & 1.1^2 & -14 & 1.2^2 & -1 \\

 f_{\pi } & .087 & .081 & -6 & .082 & -5
& .080 & -8 & .083 & -4
& .085 & -3 & .082 & -5 & .082 & -6 \\

 F_{\rho } & .15 & .12 & -23 & .16 & 6
& .18 & 22 & .14 & -8
& .12 & -17 & .14 & -4 & .15 & 1 \\

 F_{\omega } & .14 & .12 & -18 & .16 & 13
& .18 & 31 & .14 & -2
& .12 & -12 & .14 & 3 & .15 & 8 \\

 g_{\rho \pi \pi } & 6.0 & 4.8 & -20 & 5.3 & -11
& 5.5 & -8 & 5.1 & -15
& 4.7 & -21 & 5.0 & -17 & 5.1 & -15 \\

 g_{\rho \pi \gamma } & .22 & .23 & 3 & .25 & 13
& .26 & 20 & .24 & 7
& .21 & -2 & .23 & 6 & .24 & 9 \\

 g_{\omega \pi \gamma } & .72 & .68 & -6 & .74 & 3
& .79 & 10 & .71 & -2
& .64 & -10 & .70 & -3 & .72 & 0 \\

 g_{\omega \rho \pi } & 15 & 13 & -11 & 15 & 3
& 17 & 12 & 14 & -4
& 11 & -25 & 13 & -15 & 14 & -7 \\

 10^3 L_9 & 6.9 & 5.8 & -15 & 6.3 & -9
& 6.6 & -4 & 6.1 & -12
& 4.7 & -32 & 5.3 & -23 & 5.8 & -16 \\

 10^3 L_{10} & -5.5 & -5.8 & 6 & -6.3 & 15
& -6.6 & 20 & -6.1 & 11
& -4.7 & -15 & -5.3 & -3 & -5.8 & 5 \\

 r_{\pi }{}^2 & 12 & 11 & -12 & 11 & -7
& 12 & 2 & 11 & -12
& 8 & -35 & 9 & -21 & 10 & -13 \\ \hline\hline

 \multicolumn{2}{|l|}{\text{RMSE}}& \multicolumn{2}{c|}{17\%} & \multicolumn{2}{c|}{8.5\%} & \multicolumn{2}{c|}{14\%} & \multicolumn{2}{c|}{9.6\%} & \multicolumn{2}{c|}{19\%} & \multicolumn{2}{c|}{12\%} & \multicolumn{2}{c|}{8.7\%}\\ \hline\hline

 m_{\rho '}{}^2 & 1.5^2 & 2.0^2 & 87 & 1.8^2 & 37
& 1.6^2 & 15 & 1.9^2 & 58
& \text{---} & \text{---} & 1.2^2 & -31 & 1.3^2 & -24 \\

 F_{a_1} & .14 \text{ -- } .17 & .12 &  & .20 &
& .29 &  & .16 &
& .09 &  & .18 &  & .28 &  \\

 10^3 L_1 & .4 \pm .3 & .6 &  & .5 &
& .4 &  & .5 &
& .4 &  & .4 &  & .4 &  \\

 10^3 L_2 & 1.4 \pm .3 & 1.2 &  & 1.0 &
& .9 &  & 1.1 &
& .8 &  & .8 &  & .9 &  \\

 10^3 L_3 & -3.5 \pm 1.1 & -3.5 &  & -2.9 &
& -2.7 &  & -3.2 &
& -2.5 &  & -2.5 &  & -2.6 &  \\ \hline

\end{array}
\]
\caption{
All the dimensionfull quantities are in powers of \small{GeV} and the deviations are in percent.\newline
The fitted parameters (MeV): for flat space, $L^{-1} \approx 430$ and $g_5^{-2} \approx 7.6$; for AdS space, $L^{-1} \approx 320$ and $g_5^{-2} \approx 5.3$; for the Sakai-Sugimoto model, $L^{-1} \approx 260$ and $g_5^{-2} \approx 5.4$; for $\sqrt{L/z}$, $L^{-1} \approx 370$ and $g_5^{-2} \approx 6.3$;\newline for $K=2$, $f_1\approx180$, $f_2\approx120$ and $g_1\approx8.8$; for $K=3$, $f_1\approx270$, $f_2\approx130$, $g_1\approx7.3$ and $g_2\approx10$; for $K=4$, $f_1\approx460$, $f_2\approx150$, $f_3\approx140$, $g_1\approx5.3$ and $g_2\approx11$.
}
\label{table}
\end{center}
\end{table}
\end{landscape}


\end{document}